\documentclass[conference]{IEEEtran}

\usepackage{cite}
\usepackage{easybmat}
\usepackage{stfloats}
\usepackage{amsmath,amssymb,amsfonts}
\usepackage{graphicx}
\usepackage{textcomp}
\usepackage[linesnumbered,lined,boxed,commentsnumbered, ruled]{algorithm2e}
\usepackage{epstopdf}
\usepackage{balance}
\usepackage{caption}
\usepackage{subcaption}
\usepackage{algorithmic}
\usepackage{xcolor}
\usepackage[letterpaper, left=0.7in, right=0.7in, bottom=1in, top=0.7in]{geometry}
\usepackage{enumitem}

\newtheorem{prop}{Proposition}

\newenvironment{proof}{\paragraph*{Proof}}{\hfill$\square$}

\renewcommand{\baselinestretch}{.99} 
\newcommand{\mrm}[1]{\ensuremath{\mathrm{#1}}}
\renewcommand{\vec}{\ensuremath{\mrm{vec}}}
\newcommand{\diag}{\ensuremath{\mrm{diag}}}
\newcommand{\bs}[1]{\ensuremath{\boldsymbol{#1}}}
\newcommand{\comment}[1]{}

\renewcommand{\H}{\boldsymbol{H}}

\IEEEoverridecommandlockouts

\begin{document}

\title{Increased Multiplexing Gain with Reconfigurable Surfaces: Simultaneous Channel Orthogonalization and Information Embedding}
\author{Juan Vidal Alegr\'{i}a\IEEEauthorrefmark{1}, Joao Vieira\IEEEauthorrefmark{2}, Fredrik Rusek\IEEEauthorrefmark{1}\IEEEauthorrefmark{3} \\
\IEEEauthorblockA{\IEEEauthorrefmark{1}Department of Electrical and Information Technology, Lund University, Lund, Sweden\\} 
\IEEEauthorblockA{\IEEEauthorrefmark{2}Ericsson Research, Lund, Sweden\\} 
\IEEEauthorblockA{\IEEEauthorrefmark{3}Sony Research Center, Lund, Sweden\\} 
\{juan.vidal\_alegria, fredrik.rusek\}@eit.lth.se, joao.vieira@ericsson.com

\thanks{This work has been financially supported by Ericsson AB, Lund, Sweden.}}

\maketitle

\begin{abstract}
Reconfigurable surface (RS) has been shown to be an effective solution for improving wireless communication links in general multi-user multiple-input multiple-output (MU-MIMO) setting. Current research efforts have been largely directed towards the study of reconfigurable intelligent surface (RIS), which corresponds to an RS made of passive reconfigurable elements with only phase shifting capabilities. RIS constitutes a cost- and energy- efficient solution for increased beamforming gain since it allows to generate constructive interference towards desired directions, e.g., towards a base station (BS). However, in many situations, multiplexing gain may have greater impact on the achievable transmission rates and number of simultaneously connected devices, while RIS has only been able to achieve minor improvements in this aspect. Recent work has proposed the use of alternative RS technologies, namely amplitude-reconfigurable intelligent surface (ARIS) and fully-reconfigurable intelligent surface (FRIS), to achieve perfect orthogonalization of MU-MIMO channels, thus allowing for maximum multiplexing gain at reduced complexity. In this work we consider the use of ARIS and FRIS for simultaneously orthogonalizing a MU-MIMO channel, while embedding extra information in the orthogonalized channel. We show that the resulting achievable rates allow for full exploitation of the degrees of freedom in a MU-MIMO system with excess of BS antennas.
\end{abstract}


\section{Introduction}\label{sec:intro}
Massive multiple-input multiple-output (massive MIMO) \cite{marzetta} constitutes one of the main state-of-the-art solutions for base station (BS) implementation in 5G networks and beyond\cite{emil_next, 5g_book}. This technology corresponds to an evolution of the traditional multi-user MIMO (MU-MIMO) \cite{jindal} where the number of antennas at the BS grows large, leading to increased spatial resolution for effectively multiplexing user equipments (UEs) in the spatial domain \cite{marzetta,rusek}. 

Commercial deployments of massive MIMO are already available \cite{emil_next}, confirming the potential and maturity of this technology. Thus, the research community is now directing efforts towards exploring new disruptive technologies beyond massive MIMO. These technologies include large intelligent surface (LIS) \cite{husha_data}, which correspond to the natural evolution of massive MIMO by considering BSs consisting of whole walls covered with electromagnetially active material,\footnote{As discussed in \cite{husha_data}, practical implementations of LIS may consist of discrete antenna arrays, i.e., giving a sampled version of the continuous LIS.} or reconfigurable intelligent surfaces (RIS) \cite{al_zappone,survey_ris}, which consider surfaces made of passive elements with tunable reflecting coefficients, thus offering some control over propagation channel between two communication ends. This work considers some reconfigurable surface (RS) technologies, initially proposed in \cite{globecom2022}, which lie in between the concepts of LIS and RIS, but which are mainly inspired by RIS since they are used to adjust the propagation channel, i.e., they are not a BS technology like LIS.

Also known as intelligent reflecting surface (IRS), RIS constitutes an attractive enabling technology for 6G \cite{6g} due to it cost- and energy-efficient implementation, which can still lead to important improvements in link connectivity \cite{al_zappone,survey_ris}. Much of the work on RIS rely on exploiting its impressive power scaling laws \cite{survey_ris}, which is a result of the increased beamforming gain associated to redirecting the reflected waves towards the intended directions. However, RIS has also been considered for improving multiplexing capabilities either directly, e.g., by improving the rank of the single-user MIMO system \cite{rank_impr}, or indirectly, e.g., by maximizing sum-rates in different settings \cite{sum_rate,rate}. Moreover, in \cite{globecom2022}, two alternative RS technologies, namely amplitude-reconfigurable intelligent surface (ARIS) and fully-reconfigurable intelligent surface (FRIS), are considered in a MU-MIMO scenario to achieve perfect channel orthogonalization without the need for RS amplification, i.e., leading to perfect multiplexing of UEs at the BS with reduced complexity. This work builds upon the results from \cite{globecom2022}, since we use the proposed RS technologies to achieve perfect channel orthogonalization while embedding information in the process.

The use of RIS as a low-complexity energy-efficient transmitter has also been considered in various works, with interesting proposals on how the RIS can modulate information \cite{survey_ris}. These proposals combine the concept of symbiotic radio \cite{symb_radio}, where backscatter devices passively modulate information on incoming waves, with the RIS paradigm. For example, in \cite{basar_im}, index-modulation is proposed for embedding information from the RIS to both modulated and unmodulated carrier signals, while in \cite{ris_8psk} an 8-PSK modulation is practically implemented using RIS. 

On the other hand, it is also possible to use the RIS for simultaneously improving a communication link while embedding its own information. Some examples include \cite{bf_plus_ris_data}, where a solution is proposed for employing RIS to achieve beamforming gains to a UE while transmitting information to the BS by turning on/off some of its elements---thus sacrificing RIS beamforming performance---or \cite{cap_symb_ris}, which studies the capacity of a RIS-assisted single-user MIMO communication where the RIS selects its reflecting states from a predesigned codebook to embed information. However, the literature has failed to characterize the achievable multiplexing gain when using RS technologies for allowing effective multiplexing of UEs while simultaneously embedding information in the channels. To this end, we consider the RS technologies proposed in \cite{globecom2022}, ARIS and FRIS, to study an extra RS-to-BS communication link simultaneous to the UEs-to-BS over a MU-MIMO channel orthogonalized by the RS. We show that we can achieve maximum multiplexing gains, i.e., scaling with the total number of BS antennas, even when there is a large excess of BS antennas with respect to UEs.

The rest of the paper is organized as follows. Section~\ref{sec:model} presents the system model, together with some background on how to achieve channel orthogonalization with ARIS and FRIS. In Section~\ref{sec:rates} we derive the achievable rates for simultaneous UE-BS and RS-BS communication with orthogonalized channels. Section~\ref{sec:mux_gains} presents the derivation of the multiplexing gain associated to the achievable rates. Finally, Section~\ref{section:conc} concludes the paper.

% Talk about RIS solutions for Symbiotic Radio

\section{System model}\label{sec:model}
Let us consider an uplink MU-MIMO scenario where $K$ single-antenna UEs are transmitting to an $M$-antenna BS, with $M>K$, through a narrow-band channel with the aid of an RS with $N$ reconfigurable elements, $N\gg M$. The $M\times 1$ received complex vector, $\boldsymbol{y}$, can be expressed as
\begin{equation}\label{eq:ul_model}
\bs{y} = \H \bs{s} + \bs{n},
\end{equation}
where $\bs{H}$ is the $M\times K$ channel matrix, $\boldsymbol{s}$ is the $K \times 1$ vector of symbols transmitted by the users, and $\boldsymbol{n}$ is a zero-mean complex white Gaussian noise vector with sample variance $N_0$.\footnote{We assume that the RS does not add correlated noise since it may not require relevant amplification \cite{globecom2022}.} Considering that there exists a direct channel as well as a reflected channel through the RS, we can express the channel matrix as
\begin{equation}\label{eq:channel}
\H = \H_0+\H_1 \bs{\Theta} \H_2,
\end{equation}
where $\H_0$ corresponds to the $M\times K$ direct channel from the BS to the UEs, $\H_1$ and $\H_2$ correspond to the $M\times N$ channel from the BS to the RS and the $N \times K$ channel from the RS to the UEs, respectively, and $\bs{\Theta}$ is the matrix of reflection coefficients at the RS. 
\subsection{Background}
We consider two types of RS systems proposed in \cite{globecom2022}, namely ARIS and FRIS, which give the reflection matrices
\begin{equation}
    \bs{\Theta}_\mrm{ARIS} = \diag\left(\alpha_1,\dots,\alpha_N\right),\;\; \alpha_i\in \mathbb{C} \;\; \forall i,
\end{equation}
\begin{equation}
    \bs{\Theta}_\mrm{FRIS}\in \mathbb{C}^{N\times N}.
\end{equation} 
 Note the increased requirement in processing capabilities with respect to the widely studied RIS, which is typically modeled as an ARIS with the additional restriction $|\alpha_i|^2=1$ $\forall i$.
 
 In \cite{globecom2022} it is shown that both ARIS and FRIS can create a perfectly orthogonal channel, which is given by \begin{equation}\label{eq:H_U}
 \H=\sqrt{\beta}\widetilde{\bs{U}},
 \end{equation}
 where we restrict $\widetilde{\bs{U}}^\mathrm{H}\widetilde{\bs{U}}=\mathbf{I}_K$. Equivalently, we can express
 \begin{equation}\label{eq:Ut_U}
\widetilde{\bs{U}}=\bs{U}\begin{bmatrix}\mathbf{I}_K \\
\mathbf{0}_{(M-K)\times K} \end{bmatrix},
 \end{equation}
where $\bs{U}$ is an $M \times M$ unitary matrix. Note that we do not lose generality in \eqref{eq:Ut_U} by disregarding the multiplication from the right of another unitary matrix since said matrix could be absorbed by the first $K$ rows of $\bs{U}$, leading to another unitary matrix.

From \cite{globecom2022}, we can obtain the desired orthogonal channel matrix $\H=\sqrt{\beta}\widetilde{\bs{U}}$ by selecting the reflection coefficients of the ARIS as
 \begin{equation}\label{eq:ARIS_conf}
    \bs{\alpha} =  \bs{\mathcal{H}}_{12}^\dagger\mrm{vec}\left(\sqrt{\beta}\widetilde{\bs{U}}-\H_0\right),
\end{equation}
where $\bs{\mathcal{H}}_{12}^\dagger$ corresponds to the right pseudo-inverse of matrix $\bs{\mathcal{H}}_{12}=\begin{bmatrix}\vec(\bs{h}_{11}\bs{h}_{21}^\mrm{T}) & \dots & \vec(\bs{h}_{1N}\bs{h}_{2N}^\mrm{T})\end{bmatrix}$ of dimensions $MK \times N$. The existence of $\bs{\mathcal{H}}_{12}^\dagger$ gives the conditions $N>MK$, together with full-rank $\bs{\mathcal{H}}_{12}$, for appropiate operation of the ARIS. For the FRIS, the same channel would be achieved by selecting the reflection matrix as
\begin{equation}\label{eq:FRIS_conf}
    \bs{\Theta}_\mrm{FRIS}=\H_1^\dagger \left(\sqrt{\beta}\widetilde{\bs{U}}-\H_0 \right)\H_2^\dagger,
\end{equation} 
 where $\H_1^\dagger$ is the right pseudo-inverse of $\H_1$ and $\H_2^\dagger$ is the left pseudo-inverse of $\H_2$. The existence of $\H_1^\dagger$ and $\H_2^\dagger$ gives the conditions $N>\min(M,K)$, together with full-rank $\H_1$ and $\H_2$, for appropriate operation of the FRIS. Note that we can also achieve any arbitrary channel using ARIS or FRIS by substituting $\sqrt{\beta}\widetilde{\bs{U}}$ in \eqref{eq:ARIS_conf} or \eqref{eq:FRIS_conf} with the desired channel matrix, respectively. However, this work restricts to the case where the desired channel is given by \eqref{eq:H_U} due to the beneficial properties of orthogonal channels in MU-MIMO \cite{globecom2022,mimo}.

\section{Achievable rates for simultaneous RS-plus-UEs transmission}\label{sec:rates}
In Section~\ref{sec:model}, we show that ARIS and FRIS can be configured so as to generate arbitrary channel matrices. Let us, however, maintain the channel orthogonality constraint, i.e., $\H$ is restricted to \eqref{eq:H_U} with $\widetilde{\bs{U}}$ given in \eqref{eq:Ut_U}, since this corresponds to the most desirable channel structure for spatially multiplexing the UEs at reduced complexity, as discussed in \cite{globecom2022}. Note that we still have freedom in selecting $\bs{U}$ as long as it is unitary. We propose to use this freedom for embedding extra information at the RS, hence opening a new communication link between the RS and the BS which comes at essentially no cost.

In order to understand the potential of embedding extra information in the RS, we will compute $\mathcal{I}\big(\bs{y};\bs{s}, \widetilde{\bs{U}}\big)$, i.e., the mutual information between the received vector, and the RS and UEs symbols for some input distribution. This corresponds to the rate at which the RS and the UEs can simultaneously transmit information to the BS over an orthogonalized channel. However, assuming that there is no cooperation between the UEs and the RS, the information that can be transmitted by the UEs is upper bounded by $\mathcal{I}\big(\bs{y};\bs{s}|\widetilde{\bs{U}}\big)$, i.e., the mutual information of the corresponding orthogonal MIMO channel with perfect-channel state information (CSI), so $\mathcal{I}\big(\bs{y};\bs{s}, \widetilde{\bs{U}}\big)$ would correspond the achievable sum-rate for the UEs-plus-RS data. From the chain rule of mutual information we have
\begin{equation}\label{eq:sum_rates}
\mathcal{I}\big(\bs{y};\bs{s}, \widetilde{\bs{U}}\big) = \mathcal{I}\big(\bs{y};\widetilde{\bs{U}}\big)+\mathcal{I}\big(\bs{y};\bs{s}|\widetilde{\bs{U}}\big),
\end{equation}
so we can always have a non-negative information gain, given by $\mathcal{I}\big(\bs{y};\widetilde{\bs{U}}\big)$, with respect to the baseline for common MIMO systems, corresponding to $\mathcal{I}\big(\bs{y};\bs{s}|\widetilde{\bs{U}}\big)$. An interpretation of \eqref{eq:sum_rates} is that we can embed information at the RS by making use of the freedom to select $\widetilde{\bs{U}}$, which can then be extracted from $\bs{y}$ with arbitrarily small error as long as the respective information rate is below $\mathcal{I}\big(\bs{y};\widetilde{\bs{U}}\big)$ \cite{inf_th}.\footnote{Changing the variable order in the chain rule from \eqref{eq:sum_rates} gives $\mathcal{I}\big(\bs{y};\bs{s}, \widetilde{\bs{U}}\big) = \mathcal{I}\big(\bs{y};\boldsymbol{s}\big)+\mathcal{I}\big(\bs{y};\widetilde{\bs{U}} |\bs{s}\big)$, which means that we could sacrifice UE rate to achieve maximum RS rate, $\mathcal{I}\big(\bs{y};\widetilde{\bs{U}} |\bs{s}\big)$. However, this has lower practicality since the UEs are likely have more information to transfer.} Assuming the UEs transmit Gaussian symbols, $\bs{s}\sim \mathcal{CN}(\bs{0}_{K\times 1}, E_s \mathbf{I}_K)$, which corresponds to the input distribution achieving capacity for the perfect-CSI case, $\mathcal{I}\big(\bs{y};\bs{s}|\widetilde{\bs{U}}\big)$ is maximized and leads to the famous log-det formula \cite{mimo}, which for the orthogonal channel is given by
\begin{equation}\label{eq:ue_sumrate}
    \mathcal{I}\big(\bs{y};\bs{s}|\widetilde{\bs{U}}\big) = K \log\left(1+\frac{\beta E_s}{N_0}\right).
\end{equation}
It would then remain to compute $\mathcal{I}\big(\bs{y};\widetilde{\bs{U}}\big)$, which can be expressed as
\begin{equation}\label{eq:I_y_U}
    \mathcal{I}\big(\bs{y};\widetilde{\bs{U}}\big) = \mathfrak{h}(\bs{y})- \mathfrak{h}\big(\bs{y}|\widetilde{\bs{U}}\big).
\end{equation}
The conditional differential entropy $\mathfrak{h}\big(\bs{y}|\widetilde{\bs{U}}\big)$ is well defined since we have $\bs{y}|\widetilde{\bs{U}}\sim \mathcal{CN}\big(\bs{0}_{M\times 1}, \beta E_s\widetilde{\bs{U}}\widetilde{\bs{U}}^\mrm{H}+N_0\mathbf{I}_M\big)$, which gives
\begin{equation}\label{eq:h_y_U}
\begin{aligned}
    \mathfrak{h}\big(\bs{y}|\widetilde{\bs{U}}\big) =&\log \det \left(\pi \exp(1) \big( \beta E_s\widetilde{\bs{U}}\widetilde{\bs{U}}^\mrm{H}+N_0\mathbf{I}_M \big) \right)\\
    =&K\log\left(\pi \exp(1) (\beta E_s+N_0)\right)\\
    &+(M-K)\log\left(\pi \exp(1) N_0\right),
\end{aligned}
\end{equation}
where we have used \eqref{eq:Ut_U} and extracted the unitary matrix from the determinant to reach the final simplified expression. On the other hand, $\mathfrak{h}(\bs{y})$ is given by
\begin{equation}\label{eq:hy}
    \mathfrak{h}(\bs{y}) = -\mathbb{E}\{\log(p(\bs{y}))\},
\end{equation}
which may be computed through Monte-Carlo simulations by averaging over random realizations of $\boldsymbol{y}$. However, in order to compute \eqref{eq:hy} we first need to specify an input distribution for $\widetilde{\bs{U}}$ and then characterize $p(\bs{y})$, the corresponding probability distribution function (PDF) of $\bs{y}$ from said input distribution of $\widetilde{\bs{U}}$. The most meaningful input distribution for $\widetilde{\bs{U}}$ is for it to be isotropically distributed in the unitary subspace where it lies, i.e., $\widetilde{\bs{U}}$ would be constructed by \eqref{eq:Ut_U} with $\bs{U}$ uniformly distributed in the unitary space $\mathcal{U}(M)$. Furthermore, works like \cite{unitary_cap} motivate the use of isotropically distributed random matrices since they achieve capacity under Rayleigh fading scenarios, while our work considers the multiplication of the information transmitting orthogonal channel with a Gaussian vector, which is the vector equivalent of a Rayleigh fading channel. The following proposition gives the expression for $p(\bs{y})$ with isotropically distributed $\widetilde{\boldsymbol{U}}$.
\begin{prop}\label{prop:p_y}
Let $\bs{y}$ be the $M\times 1$ random vector from \eqref{eq:ul_model}, with $\boldsymbol{s} \sim \mathcal{CN}(\boldsymbol{0}_{K\times 1},E_s\mathbf{I}_K)$, and $\boldsymbol{n} \sim \mathcal{CN}(\boldsymbol{0}_{M\times 1},N_0\mathbf{I}_M)$. Assume $\bs{H}$ is given by \eqref{eq:H_U}, where $\widetilde{\bs{U}}$ is defined in \eqref{eq:Ut_U} for an isotropically distributed random unitary matrix $\bs{U}$. We can then express the PDF of $\bs{y}$ as
\begin{equation}\label{eq:p_y}
\begin{aligned}
    p(\bs{y})=\exp\left(-\frac{\Vert\bs{y}\Vert^2}{N_0}\right)\frac{(M-1)!}{(-1)^{K(M-K)}\big(\pi(\beta E_s+N_0)\big)^K}&\\
    \times \frac{\det( \bs{Z})}{(\pi N_0)^{M-K}(-\gamma\Vert\bs{y}\Vert^2)^{M-1}\;\;\prod\limits_{k=1}^{K-1}k! \prod\limits_{n=1}^{M-K-1}n!}&,
\end{aligned}
\end{equation}
where $\gamma=\frac{\beta E_s}{N_0(\beta E_s+N_0)}$, and $\bs{Z}$ is an $M \times M$ matrix whose $(i,j)$th entry is given by
\begin{equation}\label{eq:Z_ij}
    [\bs{Z}]_{i,j} =\left\{\begin{array}{lr} (\gamma\Vert\bs{y}\Vert^2)^{j-1} \exp(\gamma\Vert\bs{y}\Vert^2), \hspace{-1.5em}& j \leq K, \; i=1 \vspace{0.2em}\\
    (\gamma\Vert\bs{y}\Vert^2)^{\tilde{j}-1}, & K< j \leq M, \; i=1 \vspace{0.5em}\\
    \tfrac{\left(\tilde{i}-1\right)!}{\left(\tilde{i}-j\right)!}, & j\leq K, \; \tilde{i}\geq j \vspace{0.5em} \\
    (\tilde{i}-1)!, & K<j\leq M, \; \tilde{i} = \tilde{j}\vspace{0.5em} \\
    0, & \mrm{otherwise}, \end{array} \right.
\end{equation}
with $\tilde{j}=j-K$, and $\tilde{i}=i-1$.
\begin{proof}
See Appendix A
\end{proof}
\end{prop}

Using Proposition~\ref{prop:p_y} we can substitute $p(\bs{y})$, given by \eqref{eq:p_y}, into \eqref{eq:hy}, and compute the expected value through Monte-Carlo simulations to obtain $\mathfrak{h}(\bs{y})$. This way we can characterize $\mathcal{I}\big(\bs{y};\widetilde{\bs{U}}\big)$, hence characterizing the potential of simultaneous RS-BS and UEs-to-BS communication through orthogonalized MU-MIMO channels.

\section{Increased multiplexing gain}\label{sec:mux_gains}
In the previous section we reached a closed-form expression for $p(\boldsymbol{y})$, i.e., the PDF of the received vector $\boldsymbol{y}$ for an orthogonal channel with isotropically distributed $\widetilde{\boldsymbol{U}}$. The obtained $p(\boldsymbol{y})$ is fairly complex, so finding a closed-form expression for $\mathcal{I}\big(\bs{y};\bs{s}, \widetilde{\bs{U}}\big)$ becomes extremely challenging. However, we may simplify this expression by considering the asymptotic regime, which may allow the characterization of the multiplexing gain achieved by such a system. Let us thus focus on the high-SNR regime, i.e., $\frac{E_s}{N_0}\to \infty$. The following proposition gives the multiplexing gain associated to \eqref{eq:sum_rates}.

\begin{prop}\label{prop:mux_gain}
Let us have the same input distribution assumptions as in Proposition~\ref{prop:p_y} such that $\boldsymbol{y}$ is distributed according to the respective $p(\boldsymbol{y})$. The multiplexing gain associated to $\mathcal{I}\big(\bs{y};\widetilde{\bs{U}}\big)$, i.e., the asymptotic pre-log factor for $\frac{E_s}{N_0}\to \infty$, is given by $(M-K)$. Furthermore, the overall multiplexing gain associated to $\mathcal{I}\big(\bs{y};\boldsymbol{s},\widetilde{\bs{U}}\big)$ is given by $M$.
\begin{proof}
Without loss of generality, let us consider the high-SNR regime $\frac{E_s}{N_0}\to \infty$ by having fixed $E_s$ and $N_0\to 0$. We can then find the following limit
\begin{equation}\label{eq:lim_1}
\begin{aligned}
    \lim_{N_0\to 0} p(\boldsymbol{y}) = \frac{(M-1)!\Vert \boldsymbol{y}\Vert^{2(K-M)}}{(-1)^{(K+1)M-K^2-1} (\beta E_s)^K}&\\
     \times \frac{\det(\tilde{\bs{Z}})}{\pi^M \prod\limits_{k=1}^K k! \prod\limits_{n=1}^{M-K-1} n!}&,
\end{aligned}
\end{equation}
where the last $M-1$ rows of $\tilde{\bs{Z}}$ coincide with $\boldsymbol{Z}$ from \eqref{eq:Z_ij}, while $[\tilde{\bs{Z}}]_{1,j}=\delta_{jK}$, i.e., it is independent of $\Vert \boldsymbol{y}\Vert ^2$. Thus, we can use it in \eqref{eq:hy} to find the limit
\begin{equation}\label{eq:lim_h_y}
    \lim_{N_0\to 0} \mathfrak{h}(\bs{y}) = (M-K)\log(\beta E_s)+K \log (\beta E_s)+c_1,
\end{equation}
where $c_1$ is a constant independent of $E_s$, and thus of SNR. We can then plug \eqref{eq:lim_h_y} into \eqref{eq:I_y_U}, which using \eqref{eq:h_y_U}  leads to
\begin{equation}\label{eq:lim_I_y_U}
    \mathcal{I}\big(\bs{y};\widetilde{\bs{U}}\big)\Big|_{N_0\to 0} = (M-K)\log\Big(\frac{\beta E_s}{N_0}+c_2\Big)+c_3,
\end{equation}
where $c_2$ and $c_3$ are fixed constants independent of the SNR. The pre-log factor in \eqref{eq:lim_I_y_U} gives the multiplexing gain achieved by the RS-to-BS data transmission. By summing it to the pre-log factor $K$ from \eqref{eq:ue_sumrate}, i.e., considering \eqref{eq:sum_rates}, we obtain the overall multiplexing gain $M$. A more detailed proof may be included in the extended version, e.g., with extra steps to reach expressions like \eqref{eq:lim_1}.
\end{proof}
\end{prop}

Proposition~\ref{prop:mux_gain} shows that the considered RSs can exploit the available degrees of freedom for transmitting to the BS while a simultaneous UEs-to-BS communication is established through a desirable orthogonal channel. Furthermore, the total multiplexing gain of such a communication scheme is maximum, i.e., we could potentially transmit a number of information streams equal to the total number of BS antennas, thus taking full advantage of systems with a large excess of BS antennas, e.g., massive MIMO, LIS, cell-free massive MIMO.

\section{Numerical results}\label{section:num_res}
\begin{figure*}
     \begin{subfigure}[t]{0.49\textwidth}
         \centering
         \includegraphics[scale=0.45]{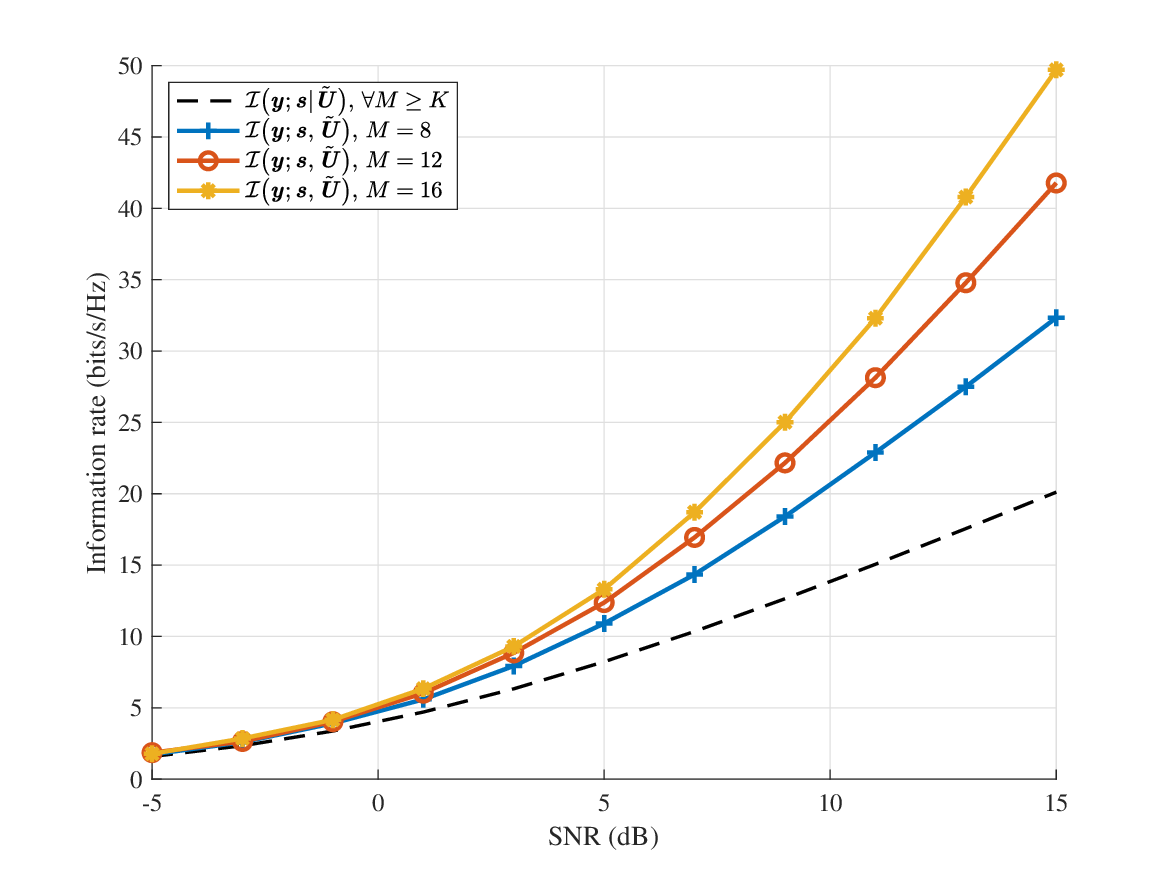}
     \end{subfigure}
     \begin{subfigure}[t]{0.49\textwidth}
         \centering
         \includegraphics[scale=0.45]{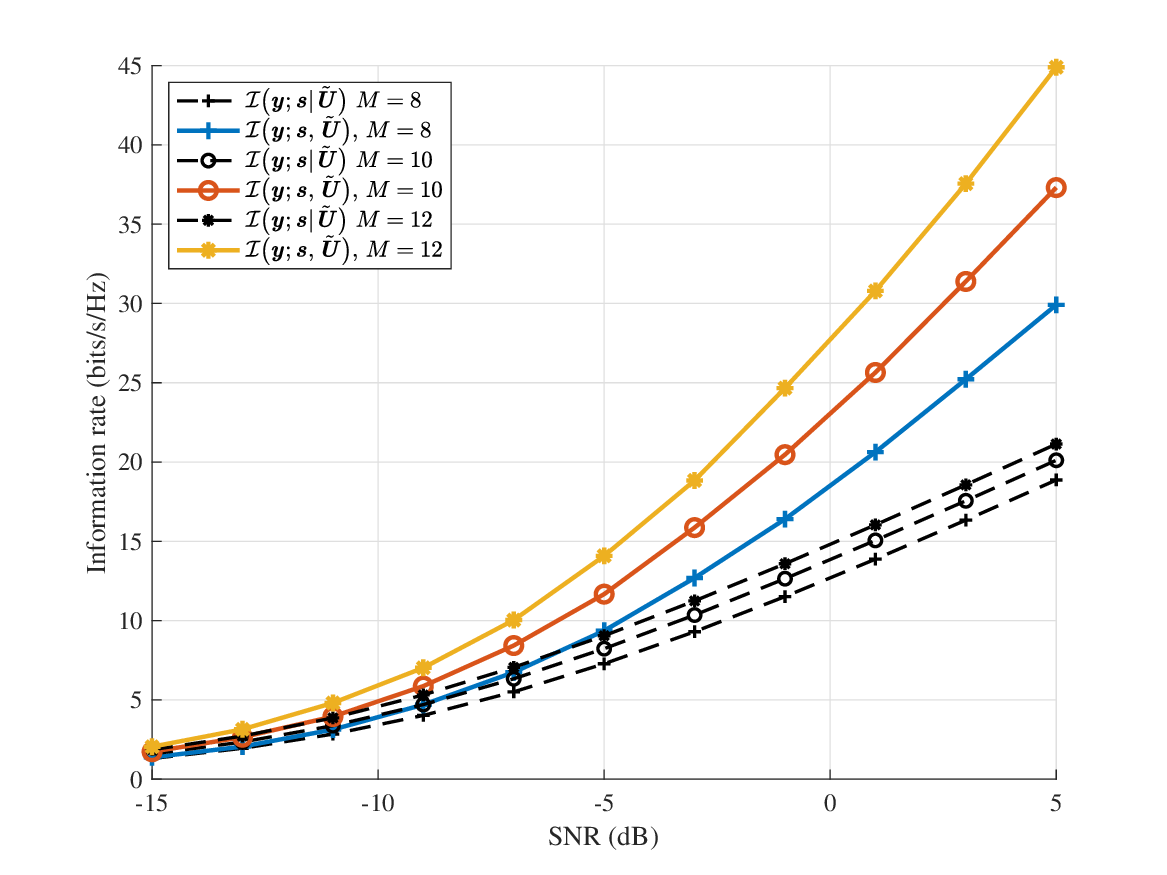}
     \end{subfigure}

     \caption{Information rates with $K=4$ UEs, and channel gain $\beta=1$ (left) and $\beta=M$ (right).}

      \label{fig:rates}
\end{figure*}

In Fig.~\ref{fig:rates} are shown the achievable UEs-plus-RS sum-rates, given by $\mathcal{I}\big(\bs{y};\bs{s},\widetilde{\bs{U}}\big)$ with isotropically distributed $\widetilde{\bs{U}}$, for different values for $M$. These rates have been computed using \eqref{eq:sum_rates}, \eqref{eq:ue_sumrate} and \eqref{eq:I_y_U}, where $\mathfrak{h}(\bs{y})$ has been computed through Mote-Carlo simulations considering \eqref{eq:hy} with $p(\boldsymbol{y})$ given by Proposition~\ref{prop:p_y}. As baseline, we may take the UEs sum rate for perfect CSI knowledge with an orthogonal channel, corresponding to $\mathcal{I}\big(\bs{y};,\bs{s}|\widetilde{\bs{U}}\big)$, i.e., given in closed-form by \eqref{eq:ue_sumrate}. Note that, since we consider perfectly orthogonal channels $\widetilde{\bs{U}}$, the UE sum rate can be shared equally among UEs without interference. In Fig.~\ref{fig:rates} (left) we ignore the possible array gains by fixing $\beta=1$, hence the lack of dependency on $M$ of $\mathcal{I}\big(\bs{y};,\bs{s}|\widetilde{\bs{U}}\big)$, while in Fig.~\ref{fig:rates} (right) we have $\beta=M$ to account for the respective array gain. Considering \eqref{eq:sum_rates}, the gap between $\mathcal{I}\big(\bs{y};\bs{s},\widetilde{\bs{U}}\big)$ and $\mathcal{I}\big(\bs{y};\bs{s}|\widetilde{\bs{U}}\big)$ corresponds to $\mathcal{I}\big(\bs{y};\widetilde{\bs{U}}\big)$, i.e., the achievable rate at which the the RS can communicate with the BS while allowing maximum transmission rate for the UEs. The results show that the extra link between the RS and the BS can exploit the $M-K$ degrees of freedom of the channel to increase the overall multiplexing gain of the transmission, which seems to scale with $M$ instead of with $K$ as for common MU-MIMO uplink transmissions \cite{mimo}. Thus, these numerical results confirm the asymptotic study from the previous section.

\section{Conclusions}\label{section:conc}
We have considered the use of two RS technologies, namely ARIS and FRIS, for simultaneous communication of UEs and RS data to a MU-MIMO BS. The main advantage of these RS technologies is that they can achieve perfectly orthogonal channels, thus allowing for perfect multiplexing of UEs in the spatial domain at reduced complexity. We have proposed employing the freedom in the desirable channel selection, which should only fulfill the orthogonality constraint, to embed extra information from the RS to the BS at essentially no cost. We have computed the mutual information of such a framework, and showed that the resulting multiplexing gain allows to exploit all the available degrees of freedom from the excess of BS antennas.

\section*{Appendix A: Proof of Proposition~\ref{prop:p_y}}
Given \eqref{eq:ul_model}, using straightforward probabilistic identities, we can express the PDF of $\bs{y}$ as
\begin{equation}\label{eq:p_y_int}
    p(\bs{y}) = \mathbb{E}_{\bs{H}} \left\{ \int_{\bs{s}\in \mathbb{C}^K} p(\bs{y}\vert \bs{H},\bs{s}) p(\bs{s}) d\bs{s} \right\}.
\end{equation}
Considering the assumptions $\bs{s}\sim\mathcal{CN}(\bs{0}_{K\times 1},E_s \mathbf{I}_K)$ and \; $\bs{n}\sim\mathcal{CN}(\bs{0}_{M\times 1},N_0 \mathbf{I}_M)$ we have
\begin{equation}\label{eq:p_s}
    p(\bs{s})=\frac{1}{(\pi E_s)^K}\exp\left(-\frac{\Vert\bs{s}\Vert^2}{E_s}\right),
\end{equation}
\begin{equation}\label{eq:p_y_Hs}
    p(\bs{y}\vert \bs{H},\bs{s})=\frac{1}{(\pi N_0)^M}\exp\left(-\frac{\Vert \bs{y}-\bs{H}\bs{s} \Vert^2}{N_0}\right),
\end{equation}
which can be substituted in \eqref{eq:p_y_int}. Given the singular value decomposition (SVD) of the channel matrix, $\bs{H}=\bs{U}\bs{\Sigma}\bs{V}^\mrm{H}$,\footnote{Note that, assuming \eqref{eq:H_U}, the SVD of $\bs{H}$ is given by \eqref{eq:Ut_U} after including the scaling of the singular values $\sqrt{\beta}$.} if we expand the norms and perform some matrix manipulations over \eqref{eq:p_y_int} (with \eqref{eq:p_s} and \eqref{eq:p_y_Hs}) we reach 
\begin{equation}
\begin{aligned}
    p(\bs{y})=\frac{\exp\left( -\frac{\Vert \bs{y}\Vert^2}{N_0}\right)}{(\pi E_s)^K(\pi N_0)^M} \quad \mathbb{E}_{\bs{H}}\Bigg\{ \int_{\tilde{\bs{s}}\in \mathbb{C}^K} \exp\left( -\frac{\Vert \tilde{\bs{s}} \Vert^2}{E_s} \right) \\
    \times \exp \left( -\frac{\tilde{\bs{s}}^\mrm{H}\bs{\Lambda}\tilde{\bs{s}}}{N_0}\right)\exp \left( \frac{2}{N_0}\Re \{ \bs{y}^\mrm{H} \bs{U} \bs{\Sigma} \tilde{\bs{s}}\} \right) d\tilde{\bs{s}}&\Bigg\},
\end{aligned}
\end{equation}
where $\bs{\Lambda}=\bs{\Sigma}^\mrm{H}\bs{\Sigma}$ is the diagonal matrix with the eigenvalues of $\bs{H}^\mrm{H}\bs{H}$, and where we have considered the change of integration variable to $\tilde{\bs{s}}=\bs{V}\bs{s}$. Further expanding the norms and multiplications leads to
\begin{equation}\label{eq:int_s}
\begin{aligned}
    p(\bs{y})= & \frac{\exp\left( -\frac{\Vert \bs{y}\Vert^2}{N_0}\right)}{(\pi E_s)^K(\pi N_0)^M} \; \mathbb{E}_{\bs{H}}\Bigg\{ \prod_{k=1}^K \int_{\tilde{s}_k\in \mathbb{C}} \exp\bigg( \\
    &   -\frac{(E_s \lambda_k+N_0) \vert \tilde{s}_k \vert^2}{E_s N_0}+\frac{2}{N_0}\Re\{\bs{y}^\mrm{H}\bs{u}_k\sigma_k \tilde{s}_k\}  \bigg)d\tilde{s}_k\Bigg\},
\end{aligned}
\end{equation}
where $\lambda_k$ is the $k$th eigenvalue of $\bs{
H}^\mrm{H}\bs{H}$, $\bs{u}_k$ is the $k$th column of $\bs{U}$, and $\sigma_k=\sqrt{\lambda_k}$ is the $k$th singular value of $\bs{H}$. The integrals from \eqref{eq:int_s} can be solved in closed form by considering the real and imaginary part of each $\tilde{s}_k$ (see \cite[Entry 3.323.2]{integrals}). Thus, after integrating and performing trivial operations we get
\begin{equation}\label{eq:py_EH}
\begin{aligned}
    p(\bs{y})=  \frac{\exp\left( -\frac{\Vert \bs{y}\Vert^2}{N_0}\right)}{\pi^K(\pi N_0)^{M-K}} \; \mathbb{E}_{\bs{H}}\Bigg\{ \prod_{k=1}^K \frac{1}{\lambda_k E_s +N_0}&  \\
     \times \exp\bigg(\frac{\lambda_k E_s\Vert \bs{u}_k \bs{y}^\mrm{H}  \Vert^2}{N_0(\lambda_k E_s+N_0)}  \bigg)&\Bigg\}.
\end{aligned}
\end{equation}
Given the orthogonality constraint on the channel \eqref{eq:H_U} with \eqref{eq:Ut_U}, we can substitute the eigenvalues $\lambda_k=\beta$ for $k\leq K$, and eigenvectors $\bs{u}_k$, directly corresponding to the $k$th row of $\widetilde{\bs{U}}$ from \eqref{eq:H_U}, which is itself a sub-matrix \eqref{eq:Ut_U} of $\bs{U}$ hereby assumed to be uniformly distributed on the unitary space $\mathcal{U}(M)$. By substituting in \eqref{eq:py_EH} and regrouping we reach
\begin{equation}
\begin{aligned}
    p(\bs{y})=&  \frac{\exp\left( -\frac{\Vert \bs{y}\Vert^2}{N_0}\right)}{\big(\pi(\beta E_s +N_0)\big)^K (\pi N_0)^{M-K}} \; \\
    &\times \mathbb{E}_{\bs{U}}\Bigg\{ 
     \exp\bigg(\frac{\beta E_s\Vert \widetilde{\bs{U}} \bs{y}^\mrm{H} \Vert^2}{N_0(\beta E_s+N_0)}  \bigg)\Bigg\},
\end{aligned}
\end{equation}
which, after expanding the expectation and operating, leads to
\begin{equation}\label{eq:py_intU}
\begin{aligned}
    p(\bs{y})=&  \frac{\exp\left( -\frac{\Vert \bs{y}\Vert^2}{N_0}\right)}{\big(\pi(\beta E_s +N_0)\big)^K (\pi N_0)^{M-K}} \; \int_{\bs{U}\in \mathcal{U}(M)} p(\bs{U})\\
        &\times \exp\left(\mrm{trace} \left(\gamma \bs{y}\bs{y}^{\mrm{H}} \bs{U}\widetilde{\bs{I}}\bs{U}^\mrm{H} \right)\right) d\bs{U},%\bs{y}\bs{y}^\mrm{H} \bs{U}\widetilde{\bs{I}}\bs{U}^\mrm{H} \right)\right) d\bs{U},
\end{aligned}
\end{equation}
where $p(\bs{U})$ is the PDF of an isotropically distributed unitary matrix, and where we have defined
\begin{equation}
    \gamma = \frac{\beta E_s}{N_0(\beta E_s+N_0)}
\end{equation}
\begin{equation}
    \widetilde{\bs{I}}= \mrm{diag}\left(\begin{bmatrix}\bs{1}_{K\times 1}\\ \bs{0}_{(M-K)\times 1}\end{bmatrix}\right).
\end{equation}
The integral \eqref{eq:py_intU} has closed-form solution \cite{HarishChandra, IZ_int}, giving
\begin{equation}\label{eq:py_U}
\begin{aligned}
    p(\bs{y})=&  \frac{\exp\left( -\frac{\Vert \bs{y}\Vert^2}{N_0}\right)}{\big(\pi(\beta E_s +N_0)\big)^K (\pi N_0)^{M-K}} \frac{\prod\limits_{m=1}^{M-1}m! \det(\bs{G})}{\Delta(\gamma \bs{y}\bs{y}^\mrm{H})\Delta(\widetilde{\bs{I}})},
\end{aligned}
\end{equation}
where $\Delta(\bs{A})=\prod_{1\leq i<j\leq T}(\lambda_{j}(\bs{A})-\lambda_{i}(\bs{A}))$ corresponds to the Vandermonde determinant of the decreasing eigenvalues $\lambda_i(\bs{A})$ of some $T \times T$ positive semi-definite matrix $\bs{A}$, and where the $(i,j)$th entry of $\bs{G}$ is given by
\begin{equation}\label{eq:Gmat}
    [\bs{G}]_{i,j} = \exp\left(\lambda_i(\gamma \bs{y}\bs{y}^\mrm{H}) \lambda_j(\widetilde{\bs{I}})\right).
\end{equation}
Given the rank deficiency of $\widetilde{\bs{I}}$, which has $K$ non-zero eigenvalues $\lambda_j(\widetilde{\bs{I}})=1$ for $j=1,\dots K$, as well as of $\gamma \bs{y}\bs{y}^\mrm{H}$, which has only one non-zero eigenvalue $\lambda_1(\gamma \bs{y}\bs{y}^\mrm{H})=\gamma \Vert \bs{y}\Vert^2$, the right quotient in \eqref{eq:py_U} evaluates to an indeterminate form $0/0$. Let us define $\bs{G}$ in function form as $\bs{G}=\{g_i(a_j)\}$ for $1\leq i,j \leq M$, where $a_j=\lambda_j(\widetilde{\bs{I}})$, which leads to
\begin{equation}
g_i(x)=\exp\left(\lambda_i(\gamma \bs{y}\bs{y}^\mrm{H}) x\right).    
\end{equation}
We can then apply \cite[Lemma~2]{limit} to find the limit
%\begin{equation}
%    \lim_{\substack{\lambda_1(\widetilde{\bs{I}}),\dots,\lambda_K(\widetilde{\bs{I}})\to 1\\ \lambda_{K+1}(\widetilde{\bs{I}}),\dots,\lambda_M(\widetilde{\bs{I}})\to 0 }} \!\!\!\! \frac{\det(\bs{E})}{\Delta(\widetilde{\bs{I}})}\!=\! \frac{\det(\bs{F})}{(-1)^{K(M-K)} \prod\limits_{k=1}^{K-1}k! \prod\limits_{{n=1}}^{M-K-1}n!},
%\end{equation}
\begin{equation}
    \lim_{\substack{a_1,\dots,a_K\to 1\\ a_{K+1},\dots,a_M\to 0 }} \!\!\!\!\!\! \frac{\det(\{g_i(a_j)\})}{\Delta(\widetilde{\bs{I}})}\!=\! \frac{\det(\bs{F})}{(-1)^{K(M-K)} \prod\limits_{k=1}^{K-1}k! \prod\limits_{{n=1}}^{M-K-1}n!},
\end{equation}
where the $i$th column of $\bs{F}$ is given by
\begin{equation}\label{eq:F_i}
    [\bs{F}]_{i,:} = \left[g_i(1) \;\; \cdots \;\; g_i^{(K-1)}(1) \;\; g_i(0) \;\; \cdots \;\; g^{(M-K-1)}_i(0)\right].
\end{equation}
Note that $g_i^{(n)}(a)$ corresponds to $n$th derivative of $g_i(x)$ at $x=a$, which is given by
\begin{equation}\label{eq:g_diff}
    g_i^{(n)}(x) = \lambda_i^{n-1}(\gamma \bs{y}\bs{y}^\mrm{H})\exp\left(\lambda_i(\gamma \bs{y}\bs{y}^\mrm{H}) x\right).
\end{equation}
After substituting \eqref{eq:g_diff} into \eqref{eq:F_i}, we can also define $\bs{F}$ in function form as $\bs{F}=\{f_j(b_i)\}$ for $1\leq i,j\leq M$, where $b_i=\lambda_i(\gamma \bs{y}\bs{y}^\mrm{H}))$, leading to
\begin{equation}
    f_j(x) = \left\{\begin{array}{lr}
        x^{j-1}\exp(x),   & j\leq K \\
         x^{\tilde{j}-1},  & K< j \leq M,
    \end{array} \right.
\end{equation}
with $\tilde{j}=j-K$. We can then apply \cite[Lemma~2]{limit} again to find the limit
\begin{equation}
    \lim_{\substack{b_1 \to \gamma \Vert \bs{y} \Vert^2 \\ b_2,\dots,b_M\to 0 }} \!\!\!\! \frac{\det(\{f_j(b_i)\})}{\Delta(\gamma \bs{y}\bs{y}^\mrm{H})}\!=\! \frac{\det(\bs{Z})}{(-\gamma \Vert\bs{y}\Vert^2)^{M-1} \prod\limits_{q=1}^{M-2}q!},
\end{equation}
where the $j$th column of $\bs{Z}$ is given by
\begin{equation}\label{eq:Z_j}
    [\bs{Z}]_{:,j} = \left[f_j(\gamma \Vert\bs{y}\Vert^2) \;\; f_j(0) \;\; f_j' \;\; \cdots \;\; f_j^{(M-2)}(0)\right]^\mrm{T},
\end{equation}
with the $n$th derivative of $f_j(x)$ given by
\begin{equation}\label{eq:fj_diff}
    f_j^{(n)}(x) = \left\{\begin{array}{lr}
        \sum\limits_{k=0}^n \binom{n}{k}\prod\limits_{l=1}^{n-k}x^{j-n+k-1}\exp(x),  \hspace{-0.5em} & j\leq K \vspace{0.3em}\\
         \prod\limits_{l=1}^{n} (\tilde{j}-l)x^{\tilde{j}-n-1}\exp(x),  & K< j \leq M,
    \end{array} \right.
\end{equation}
Substituting \eqref{eq:fj_diff} in \eqref{eq:Z_j} and operating leads to \eqref{eq:Z_ij}. Furthermore, substituting \eqref{eq:py_U} with the obtained limits leads directly to \eqref{eq:p_y}, which concludes the proof.

\bibliographystyle{IEEEtran}

\bibliography{IEEEabrv,bibliography}

%okay that works
% \fi

\end{document}